\documentclass[aip,jcp,reprint]{revtex4-2}

\usepackage[english]{babel}
\usepackage{braket}
\usepackage{amsthm}

\usepackage{mathtools}
\usepackage{siunitx}
\usepackage{accents}
\usepackage{natbib}
\usepackage{xcolor}

\usepackage{graphicx}
\usepackage{dcolumn}
\usepackage{bm}

\usepackage[utf8]{inputenc}
\usepackage[T1]{fontenc}
\usepackage{mathptmx}
\usepackage{etoolbox}
\usepackage[version=4]{mhchem}

\usepackage[normalem]{ulem}

\usepackage[]{xcolor}
\usepackage{soul}
\usepackage{cancel}
\usepackage{pdfpages}

\newcommand{\revN}[1]{{\color{black}#1}}
\newcommand{\revHK}[1]{{\color{black}#1}}
\newcommand{\revE}[1]{{\color{black}#1}}
\DeclareSIUnit\angstrom{\text {Å}}
\newcommand{\Dbraket}[2]{\langle #1 \hspace{.10em} \vert \hspace{.10em}  #2 \rangle}

\makeatletter
\AtBeginDocument{\let\LS@rot\@undefined}
\makeatother

\begin{document}


\title{Understanding failures in  electronic structure methods arising from the geometric phase effect}
%


\author{Eirik F. Kjønstad}
\email[]{eirik.kjonstad@ntnu.no}
\affiliation{Department of Chemistry, Norwegian University of Science and Technology, 7491 Trondheim, Norway}

\author{Henrik Koch}
\email[]{henrik.koch@ntnu.no}
\affiliation{Department of Chemistry, Norwegian University of Science and Technology, 7491 Trondheim, Norway}

\begin{abstract}
The geometric phase effect arises from the dependence on the nuclear coordinates in the electronic Hamiltonian, leading to sign changes of the electronic wave functions upon traversal of certain paths in nuclear configuration space. The geometric phase effect can have important consequences for the electronic structure problem, but this fact has largely gone unnoticed. We show how the geometric phase effect can significantly impact the accuracy of approximate electronic structure methods. In particular, we prove that for paths that enclose a conical intersection, any component of the wave function (such as an approximation to it) must vanish exactly, unless the associated conical intersections of the component and the wave function coincide. This has implications for methods that employ intermediate normalization, where the contribution along a reference wave function is fixed. 
We demonstrate numerically that the failure to account for the phase effect leads to asymptotic discontinuities in the wave function parameters. This results in breakdowns in coupled cluster methods or perturbation theories converging to excited states rather than the ground state. The global nature of the geometric phase effect means that these failures can span extended regions of nuclear configuration space, including regions far away from any conical intersection.

\end{abstract}


\maketitle 

\section{Introduction}
\noindent

In a seminal paper from 1975, Longuet-Higgins~\citep{longuet1975intersection} presented topological arguments proving
that traversing a loop in internal coordinate space that encloses a conical intersection leads to the geometric phase effect.\citep{longuet1958studies,berry1984quantal,domcke2004conical,yarkony1996diabolical} Starting from some initial point (or molecular geometry), and moving continuously along such a closed loop, one finds that the electronic wave functions of the intersecting states have changed sign upon returning to the initial \revE{point}. 
This geometric phase effect occurs even when the path that encloses the degeneracy is nowhere near
the actual point of degeneracy; in particular, the effect is independent of the path and requires only that the path encloses 
\revE{one}
point of degeneracy. As a result, the phase effect has global consequences for the electronic wave functions.

The phase of the electronic wave function is normally thought to be a quantity that only has a direct 
\revE{relevance for}
nuclear dynamics simulations, where a consistent phase must be maintained 
\revE{for the derivative coupling elements when integrating the} nuclear Schrödinger equation.~\citep{mead1992geometric} Indeed, the phase does not enter into the electronic Scrödinger equation at all. However, this point of view overlooks the perhaps surprising indirect effects \revE{that are} caused by  
the geometric phase effect at conical intersections. As we will show here, the geometric phase effect 
\revE{is fundamentally important to the electronic structure problem,}
\revE{leading to wide-ranging failures in approximate methods.} 

This paper is organized as follows. We start by proving the \revE{main result of the paper, the} vanishing component theorem\revE{,} in Section II. \revE{In} Section III, we analyze the consequences for intermediately normalized wave function parameterizations. In Sections IV, V, and VI, we analyze an analytical model, coupled cluster theory, and Møller-Plesset perturbation theory. Finally, in Section VII we give \revE{some concluding remarks.} 

\section{The Vanishing component Theorem}

We consider
a real electronic Hamiltonian defined on a finite-dimensional Hilbert space \revE{and focus on one eigenstate (electronic wave function) of this Hamiltonian}. Let $\ket{\Phi_A}$ be a normalized approximate wave function within a subspace, and $\ket{\Phi_F}$ the corresponding full-space wave function, such that
\begin{equation}
    \ket{\Phi_F} =  \ket{\Phi_A} c_0 +  \ket{\Phi_C},
\end{equation}
where $\ket{\Phi_C}$ lies in the orthogonal complement \revE{of $\ket{\Phi_A}$, such that}  $\braket{\Phi_A | \Phi_C} = 0$.

\begin{figure}[htb]
    \centering
    \includegraphics[width=\linewidth]{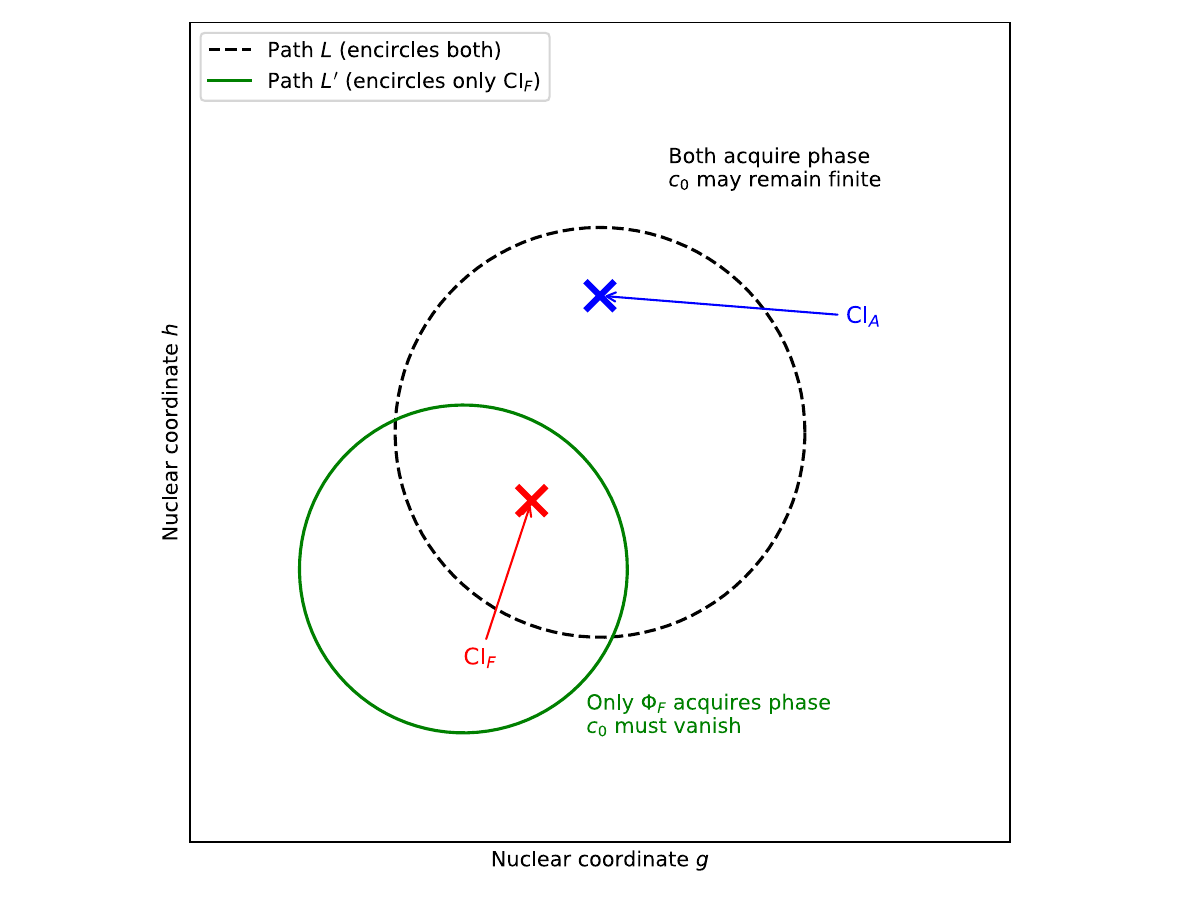}
    \caption{Illustration of the vanishing component theorem. In blue is the conical intersection (CI) associated with the subspace, and in red the CI associated with the full space. This shows that unless the two CIs coincide, it is always possible to find an alternative path (in green) where only the full space wave function acquires a phase.}
    \label{fig:proof}
\end{figure}

Now consider a closed path $L$ in nuclear configuration space, along which the full wave function $\ket{\Phi_F}$ acquires a geometric phase due to an enclosed conical intersection (\revE{see \revHK{Fig.} \ref{fig:proof}}). There are two possibilities for the behavior of $\ket{\Phi_A}$ along this path:

	1.	If $\ket{\Phi_A}$ does not acquire a geometric phase along $L$: In this case, $\ket{\Phi_A}$ remains single-valued, while $\ket{\Phi_F}$ changes sign due to the geometric phase. This implies that the coefficient $c_0$ must change sign along the loop, and therefore must vanish at some point along $L$.
    
	2.	If $\ket{\Phi_A}$ does acquire a geometric phase along $L$: This phase must arise from a conical intersection enclosed by the path, specific to the subspace in which $\ket{\Phi_A}$ resides. 

If the conical intersection associated with $\ket{\Phi_A}$ is located at a different point in nuclear configuration space than the one associated with $\ket{\Phi_F}$, then one can always construct an alternative path that encloses only the full-space conical intersection and not the subspace one. Along such a path, $\ket{\Phi_A}$ would remain single-valued, and thus $c_0$ must again vanish at some point along this new path.

Therefore, we conclude that unless the conical intersections 
\revE{of both wave functions}
coincide, there will always exist a path along which $c_0$ must pass through zero. 
Only when the two intersections are located at the same point is it impossible to isolate the geometric phase to the full wave function. 
We refer to this as the vanishing component theorem.

An immediate consequence of the theorem is that the entire crossing seam in the full space and subspace must coincide, in order to avoid regions where the subspace component vanishes. If the seams do not coincide, the dimensionality of the region where the subspace components vanish is $(N-1)$, where $N$ is the number of internal nuclear degrees of freedom. This compares to the $(N-2)$ dimensionality of the crossing seam. \revE{The dimensionalities of these subspaces are discussed further below.} 

\section{Intermediate normalization}

\begin{figure*}[htb]
    \centering
    \includegraphics[width=\linewidth]{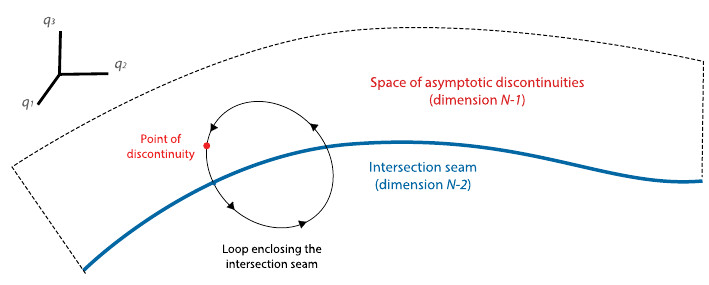}
    \caption{Asymptotically discontinuous components for intermediately normalized methods with $N = 3$ internal coordinates (labeled $q_1$, $q_2$, and $q_3$). The conical intersection seam is a curve ($N-2 = 1$) and the space of discontinuous components forms a plane ($N - 1 = 2$). From an arbitrary starting point, a loop enclosing the intersection seam will necessarily pass through a point of discontinuity where components of the wave function diverge.}
    \label{fig:discontinuity}
\end{figure*}

\revE{We next consider how the vanishing component theorem affects methods that rely on intermediate normalization.}
Here, the wave function $\ket{\Psi}$ is normalized with respect to some $\ket{\Phi}$ (often the closed-shell Hartree-Fock wave function by requiring that
\begin{align}
    \Dbraket{\Phi}{\Psi} = 1.
    \label{eq:intermediate}
\end{align}
This normalization does not imply any loss of generality as long as $\Dbraket{\Phi}{\Psi} \neq 0$. However, if $\ket{\Psi}$ has a geometric phase effect, then $\Dbraket{\Phi}{\Psi}$ will change sign continuously and pass through zero as we traverse around the conical intersection, regardless of the choice of any phase-free $\ket{\Phi}$. 
Indeed, if the wave function $\ket{\Psi}$ is to change its sign, all of its components (including the component along $\ket{\Phi}$) must go through zero at some point along the loop.
Renormalization of the wave function to enforce \revHK{Eq.(\ref{eq:intermediate})}, that is,
\begin{equation}
    \ket{\Psi'} =  \ket{\Psi}/\Dbraket{\Phi}{\Psi},
\end{equation}
will therefore lead to asymptotic discontinuities in the expansion coefficients of  
$ \ket{\Psi'}$. To see this, note that if the component along $\ket{\Phi}$ is required to be 1, then the effective weight along $\ket{\Phi}$ can only go to zero if the other components of $\ket{\Psi'}$ diverge to infinity. Moreover, the change in the sign of the component along $\ket{\Phi}$, as it passes through zero, implies a discontinuous change of sign in the diverging components. These asymptotic discontinuities (see Figure \ref{fig:discontinuity}) form a subspace of dimension $N-1$, where $N$ is the number of internal coordinates, which extends from the conical intersection seam of dimension $N-2$. 

We can therefore conclude that intermediate normalization is incompatible with a correct description of the geometric phase effect \revE{in one sense:} there is no continuous description that correctly captures the change in sign \revE{upon a complete rotation about a conical intersection}. Nevertheless, intermediate normalization has been regarded as a convenient starting point for developing many electronic structure methods.\citep{helgaker2013molecular,helgaker2012recent} However, the implications of this choice have, to the best of our knowledge, not been discussed in detail before.

\revE{For example, consider the typical case of intermediate normalization based on a}
closed-shell Hartree-Fock wave function 
\begin{equation}
    \ket{\Phi} = \ket{\mathrm{HF}} = \prod_i a^\dagger_{i\alpha}a^\dagger_{i\beta}\ket{\mathrm{vac}}.
\end{equation}
\revE{The sign of $\vert \mathrm{HF} \rangle$}
cannot be changed by an orthogonal transformation of the orbitals.
This wave function is therefore unable to account for the geometric phase effect. \revE{As a result, a correlated state $\ket{\Psi}$ based on this reference, via intermediate normalization, will exhibit asymptotic discontinuities.}

\revE{We should note that}
an open-shell determinant may change sign if one of the singly occupied orbitals acquires a geometric phase when passing through the closed loop. However, this geometric phase arises from a degeneracy among the orbital energies, and by the vanishing component theorem, we will 
\revHK{always} be able to find another path where 
\revE{the correlated state acquires a phase but the open-shell determinant does not.}

Therefore, the contribution from the open shell determinant to the exact wave function must necessarily vanish at some point along the path. The only exception is that the crossing seam in the orbital energies coincides with the seam in the correlated model, which is\revE{, in general,} highly unlikely.

\begin{figure*}[htb]
    \centering
    \includegraphics[width=\linewidth]{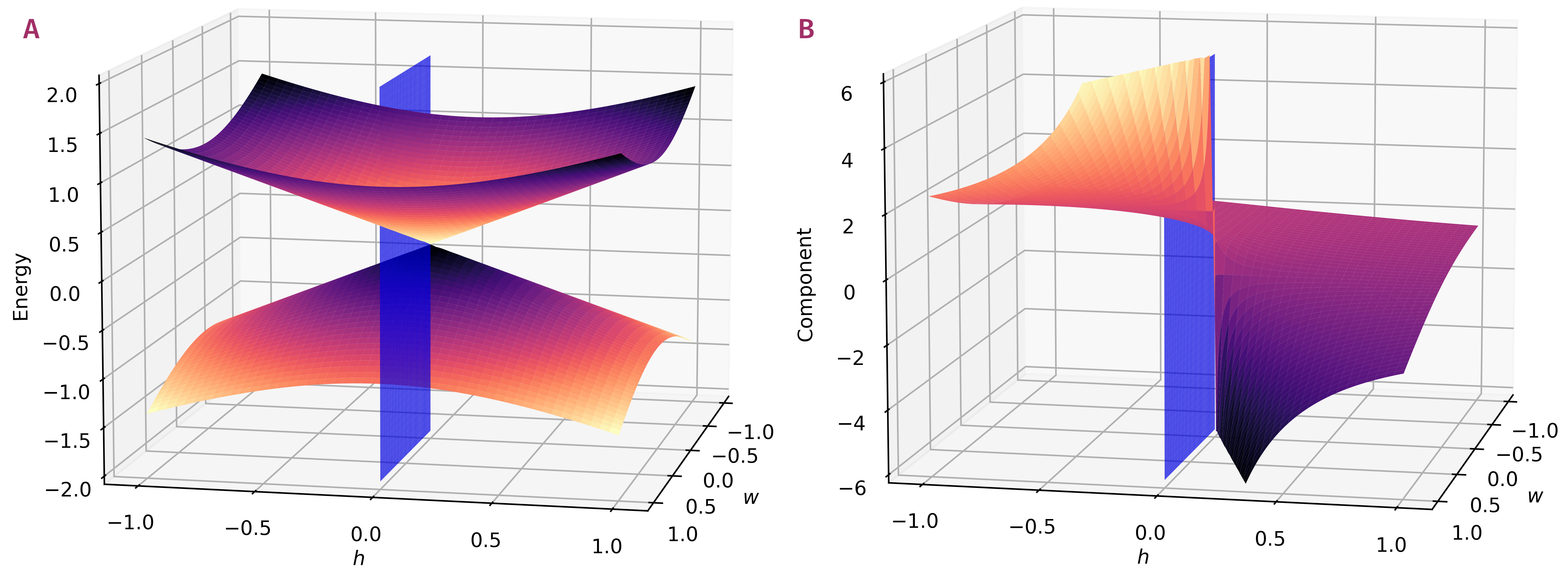}
    \caption{\textbf{A}. The energies $\lambda_\pm$ in the $(w,h)$ plane. \textbf{B}. The diverging component of the intermediately normalized ground state $\tilde{\boldsymbol{v}}_-$ in the $(w,h)$ plane. A similar divergence occurs for the excited state $\tilde{\boldsymbol{v}}_+$.  The plane where the divergence occurs is shown in blue.}
    \label{fig:divergence-analytical}
\end{figure*}

\clearpage
\section{Analytical model}
\label{Sec:analytical}
\noindent
\revE{To further characterize these asymptotic discontinuities,}
it is instructive to first consider a two-state Hamiltonian where the two states are the ground and the first excited state. In the case of a \revE{real-valued} Hamiltonian, we \revE{can} write
\begin{align}
    \boldsymbol{H} = \begin{pmatrix}
        v & 0 \\
        0 & v
    \end{pmatrix} + \begin{pmatrix}
        w & h \\ h & -w
    \end{pmatrix} = v \boldsymbol{I} + \boldsymbol{H}_c.
\end{align}
The eigenvectors of $\boldsymbol{H}$ are uniquely determined by $\boldsymbol{H}_c$ and the $v\boldsymbol{I}$ term only provides a constant shift of the energies. We will therefore restrict our analysis to $\boldsymbol{H}_c$.  Considering $\boldsymbol{H}_c$, we can express its eigenvectors and eigenvalues using polar coordinates,
\begin{equation}
\begin{aligned}
    &w = r \cos 2 \theta \\ &h = r \sin 2 \theta.
\end{aligned}
\end{equation}
Note that $\theta = 0$ to $\theta = \pi$ corresponds to a full revolution about the origin. 
The eigenvalues and eigenvectors are
    $\lambda_{\pm} = \pm r = \pm \sqrt{w^2 + h^2}$
and
\begin{equation}
\begin{aligned}
&\boldsymbol{v}_-(\theta) = \begin{pmatrix}
    - \sin \theta \\ \cos \theta
\end{pmatrix} \\
    &\boldsymbol{v}_+(\theta) = \begin{pmatrix}
        \cos \theta \\ \sin \theta
    \end{pmatrix} 
\end{aligned}
\end{equation}
respectively. 
\revE{The phase effect can now be easily investigated, since}
a path about a \revE{conical intersection} in internal coordinate space corresponds to a path about the $(w,h)$ origin.\citep{williams2023,mead1992geometric,Ruedenberg1991} Indeed, by varying $\theta$ from $0$ to $\pi$ (corresponding to a full revolution about the degeneracy), we find that the eigenvectors continuously change their sign upon returning to the same point:
\begin{align}
    \boldsymbol{v}_\pm(\pi) = -\boldsymbol{v}_\pm(0).
\end{align}
This is the geometric phase effect. Intermediate normalization changes this analysis in important ways. When
\begin{equation}
\begin{aligned}
    &\tilde{\boldsymbol{v}}_\pm(\theta) = N_\pm(\theta) \boldsymbol{v}_\pm(\theta)\\
     &N_-(\theta) = -\frac{1}{\sin \theta} \\
     &N_+(\theta) = \frac{1}{\cos \theta},
\end{aligned}
\end{equation}
we can ensure that the first component of both eigenvectors always has the same weight:
\begin{equation}
\begin{aligned}
    &\tilde{\boldsymbol{v}}_-(\theta) = \begin{pmatrix}
        1 \\ -\cot \theta
    \end{pmatrix} \\
    &\tilde{\boldsymbol{v}}_+(\theta) = \begin{pmatrix}
        1 \\ \tan \theta
    \end{pmatrix}.
\end{aligned}
\end{equation}
However, the norm of these vectors diverges at the points where the first component of the normalized vectors goes through zero, and in these points, the sign of the diverging component changes discontinuously. These asymptotic discontinuities
are shown in Fig.~\ref{fig:divergence-analytical}. We note that any path around the origin will pass through \revE{at least} one such point. \revE{Noting that the conical intersection seam is a curve for $N = 3$, the} result is a plane of asymptotic discontinuities  \revE{that extends out from the seam} (\revE{as illustrated in}  Fig.~\ref{fig:discontinuity}).

\begin{figure*}[htb]
    \centering
    \includegraphics[width=\linewidth]{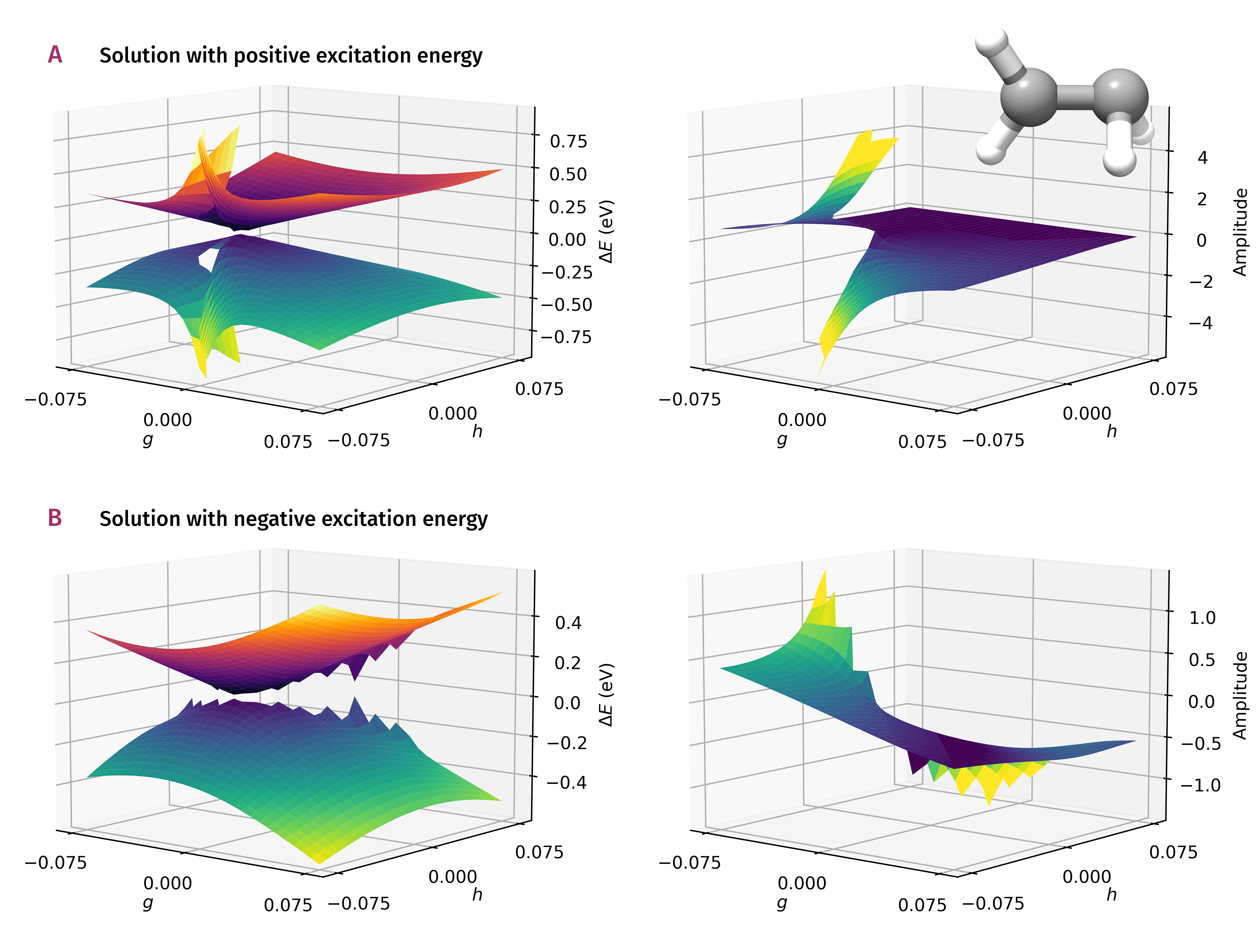}
    \caption{\textbf{A}.~Branching plane for pyramidalized ethylene $S_0$/$S_1$ minimum energy conical intersection in the case of the positive excitation energy solution of the CCSD/aug-cc-pVDZ equations.  \textbf{B}.~Same branching plane in the case of the negative excitation energy solution of the CCSD/aug-cc-pVDZ equations. The left panels in \textbf{A} and \textbf{B} show the energy difference $\Delta E$ between $S_0$ and $S_1$ and the right panels show the largest cluster amplitude in $T_1$.}
    \label{fig:ethylene-loop-cc}
\end{figure*}

\section{Coupled Cluster Theory}
\label{Sec:Reformulation of electronic structure methods to CBOA}
\noindent
\revE{We can now discuss how these discontinuities manifest in some electronic structure methods, starting with} standard coupled cluster \revE{(CC)} theory.\citep{helgaker2013molecular,bartlett2007coupled} \revE{In this method, one assumes} that the weight along the reference state, usually the closed-shell Hartree-Fock wave function $\ket{\mathrm{HF}}$, \revE{is exactly equal to} \textit{one}. \revE{In more detail, the ground state is expressed as}
\begin{align}
     \begin{split}
         \ket{\mathrm{CC}} &= \exp(T)\ket{\mathrm{HF}} \\
       &= (1 + T + \tfrac{1}{2}T^2 + \ldots)  \ket{\mathrm{HF}}, \\
     \end{split}
\end{align}
where $T = \sum_\mu t_\mu \tau_\mu$ and
\begin{align}
\Dbraket{\mathrm{HF}}{\mathrm{CC}} = 1. \label{eq:intermediate-norm-cc}
\end{align}
The $\tau_\mu$ operators in $T$ generate excitations from the reference wave function $\ket{\mathrm{HF}}$ and are usually truncated at a given excitation order (for example, $T = T_1 + T_2$ in the singles and doubles model, i.e.~CCSD\citep{purvis1982full}). 
\revE{Note that Eq.~\eqref{eq:intermediate-norm-cc} corresponds to the intermediate normalization condition, with $\ket{\Phi} = \ket{\mathrm{HF}}$ and $\ket{\Psi} = \ket{\mathrm{CC}}$.}

\revE{From this, we can immediately conclude that} the cluster amplitudes $t_\mu$ must diverge for the weight along the reference $\ket{\mathrm{HF}}$ to effectively become zero. 
Consequently, the geometric phase effect implies that the cluster amplitudes will diverge, showing an asymptotic discontinuity at a specific point along any path that encircles a ground state intersection.

\revE{As we show below, this}
leads to severe artifacts in the potential energy surfaces unless the cluster operator is complete. To understand why \revE{this occurs}, \revE{it is useful to express} $\ket{\mathrm{CC}}$ as a \revE{linear} configuration interaction expansion:
\begin{align}
    \ket{\mathrm{CC}} = (1 + T + \tfrac{1}{2}T^2 + \ldots)  \ket{\mathrm{HF}} = (1 + C) \ket{\mathrm{HF}}. \label{eq:CC-CI}
\end{align}
We may consider the operator $C$ to be a function of $T$, defining a map from $T$ to $C$. From the analysis \revE{in Section \ref{Sec:analytical} (see also Fig.~\ref{fig:divergence-analytical})}, we should expect a divergence in $C$ with opposite signs on either side of the plane with asymptotic discontinuities. If we denote the wave function on either side of the plane by $\ket{\mathrm{CC}_+}$ and $\ket{\mathrm{CC}_-}$, we thus expect that
\begin{equation}
\begin{aligned}
    \ket{\mathrm{CC}_+} &= (1 + C_+) \ket{\mathrm{HF}} =  (1+C)\ket{\mathrm{HF}} \\
    \ket{\mathrm{CC}_-} &= (1 + C_-)\ket{\mathrm{HF}} = (1-C)\ket{\mathrm{HF}}
\end{aligned}
\end{equation}
as the plane \revE{of discontinuities} is approached from either side. 
Since the coefficients in $C$ tend to infinity, these wave functions become asymptotically identical up to a sign when sufficiently close to the plane. That is,
\begin{equation}
\begin{aligned}
    \ket{\mathrm{CC}_+} &= (1+C)\ket{\mathrm{HF}} \approx C \ket{\mathrm{HF}} \\
    \ket{\mathrm{CC}_-} &= (1-C)\ket{\mathrm{HF}} \approx -C \ket{\mathrm{HF}}.
\end{aligned}
\end{equation}
\revE{At this point, a breakdown of the method can be predicted due to the following contradiction: there} is in general no pair of cluster operators ($T_+$, $T_-$) that provides the same wave function up to a sign. The map from $T$ to $C$ is not invertible for truncated $T$ and only in the complete $T$ limit (full coupled cluster or full configuration interaction limit) will there exist a $T_-$ and $T_+$ such that the wave functions become $(1 - C)\ket{\mathrm{HF}}$ and $(1 + C)\ket{\mathrm{HF}}$ \revE{with $C$ diverging to infinity}. From an equivalent perspective, the problem arises as the truncated cluster operator $T$ does not have sufficient flexibility to control the sign of the high-order excitations that result from products of cluster operators in the expansion, see eq.~\eqref{eq:CC-CI}. 
This leads to a breakdown of the method in which the potential energy surfaces can even become multi-valued upon traversal around intersections between the ground and first excited states.

\begin{figure*}[htb]
    \centering
    \includegraphics[width=\linewidth]{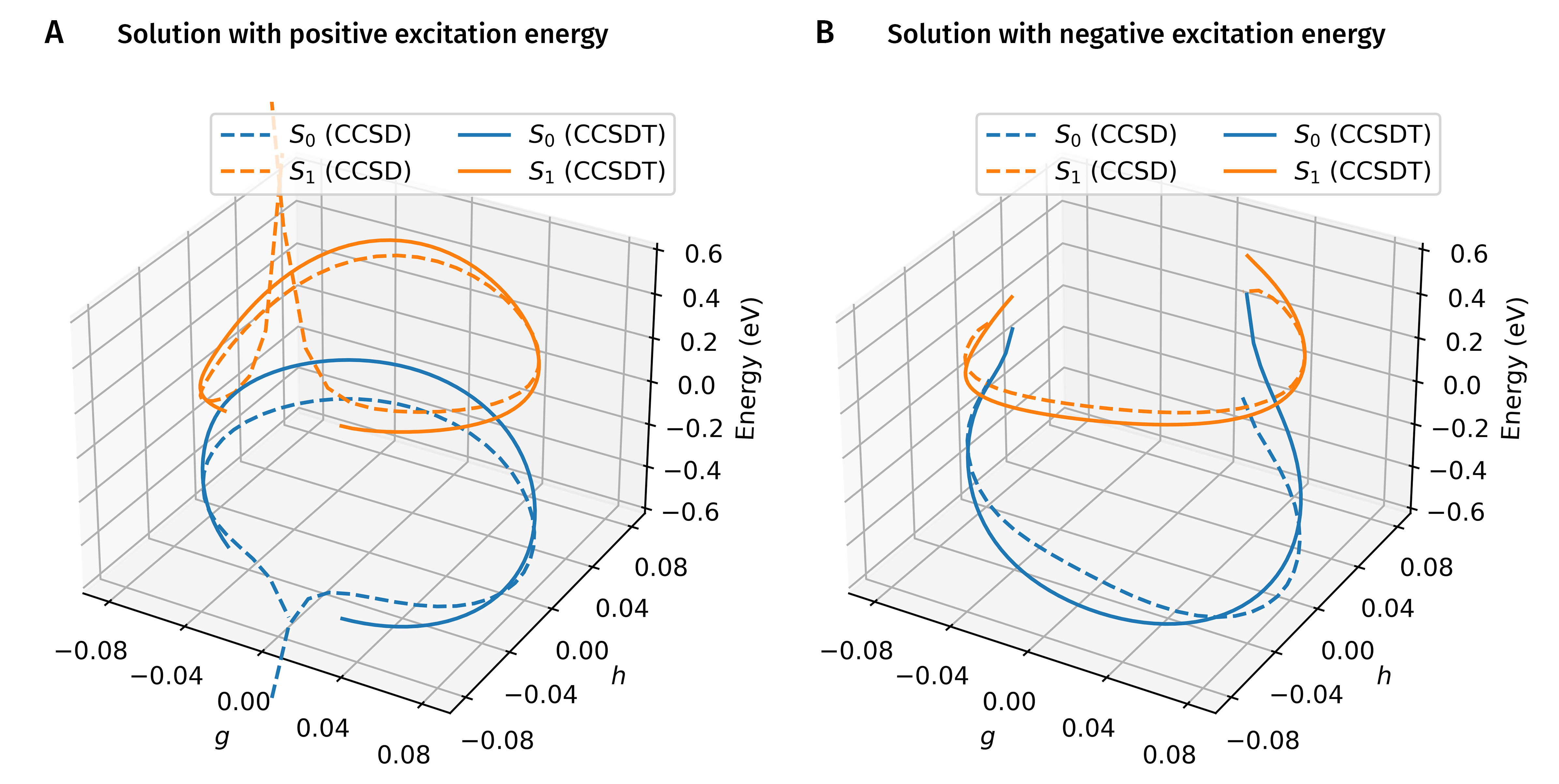}
    \caption{Circle scans in the $gh$ plane for the ethylene $S_0$/$S_1$ minimum energy conical intersection using CCSD and CCSDT with aug-cc-pVDZ.
    \textbf{A}.~ Positive excitation energy solution.  \textbf{B}.~Same scan in the case of the negative excitation energy solution. The CCSD and CCSDT energies have been shifted with respect to the average energy over the circle scan for convenient comparison of potential energy curves.
    }
    \label{fig:CCSDT}
\end{figure*}

We illustrate this behavior in the case of the pyramidalized ground state minimum energy conical intersection (MECI) in ethylene using the equation of motion CCSD method.\citep{koch1990coupled,stanton1993equation} In these calculations, we have used the CCSD minimum energy conical intersection ($\varepsilon$-MECI) structure  reported by Angelico \textit{et. al.}~\citep{Angelico_2024} in a recent work on MECI optimizations in coupled cluster theory. All calculations are performed using \revE{a development version of} the $e^T$ program,~\citep{folkestad20201} with molecular gradients and derivative couplings (which are required to obtain the $\boldsymbol{g}$ and $\boldsymbol{h}$ vectors~\citep{yarkony1996diabolical}) evaluated using recent implementations.\citep{schnack2022efficient,kjonstad2023communication} 

The potential energy surfaces in the branching plane are shown in Figure~\ref{fig:ethylene-loop-cc}.
We find two solutions, one with a positive excitation energy and one with a negative excitation energy. These two solutions correspond to the two cases where the coupled cluster state $\ket{\psi} = \exp(T)\ket{\mathrm{HF}}$ converges either to the ground state or the first excited state, respectively. Normally, the coupled cluster state is assumed to correspond to the ground state, but this is not guaranteed. The presence of multiple solutions is associated with singularities in the coupled cluster Jacobian matrix (corresponding to a vanishing excitation energy\citep{koch1990coupled,stanton1993equation}) and are therefore expected to occur at ground state intersections. The existence of such solutions has been discussed previously.\citep{piecuch2000search,sverrisdottir2024exploring} Both solutions exhibit the same behavior in the cluster amplitudes, which tend to positive and negative infinity as the plane of discontinuities is approached in the clockwise and counterclockwise directions. The breakdown covers a larger range in the case where the coupled cluster state represents the first excited state. 
Only for the positive energy solution do we obtain multi-valued potential energy surfaces, where returning to the same point does not yield the same potential energies. For the negative energy solution, the breakdown occurs more quickly, and the surfaces are therefore not multi-valued.

\revE{\revHK{Due to} the incompleteness of the cluster operator, these breakdowns will} also persist at higher levels of theory. In Fig.~\ref{fig:CCSDT}, this is shown for a scan in the $gh$ plane of the ethylene MECI with radius $r = 0.075$ using CCSD and CCSDT. Note that the region where the equations do not converge is larger for CCSDT \revE{than for CCSD}. 

The existence of multi-valued solutions is due to the inability of the cluster operator to control the sign of the wave function. \revE{Indeed, upon} a full revolution starting from a point close to the discontinuous plane, this leads to similar wave functions, up to a sign, but not identical. We emphasize the generality of this result. Based on \revE{our analysis}, this \revE{incorrect} behavior will be \revE{found} around all conical intersections between the ground and the first excited states in coupled cluster theory.

\begin{figure*}[htb]
    \centering
    \includegraphics[width=\linewidth]{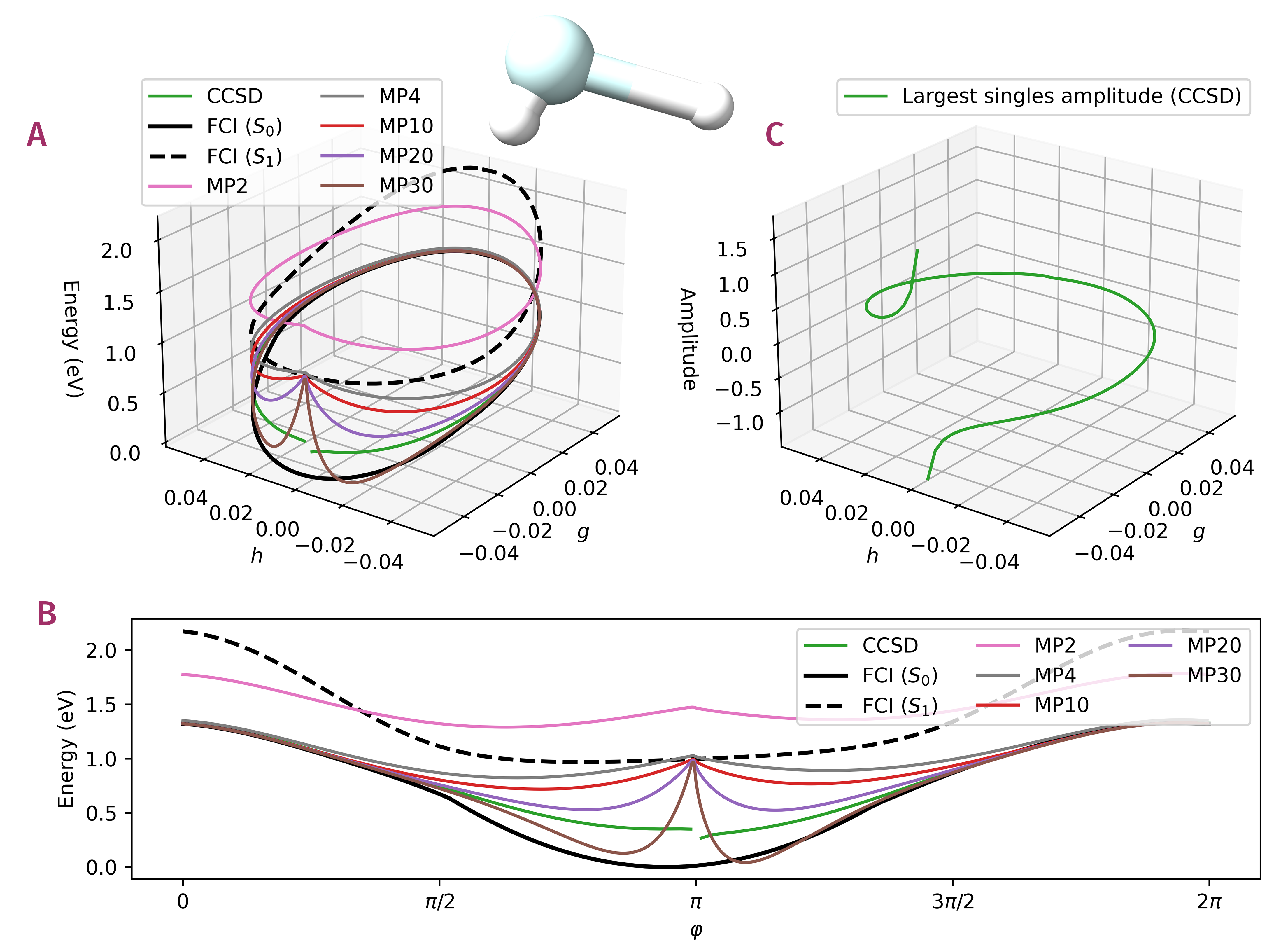}
    \caption{\textbf{A} Branching plane close to a \ce{HeH2}/aug-cc-pVDZ intersection. Shown are ground state energies given by FCI (exact), CCSD, and Møller-Plesset perturbation theory at orders 10, 20, and 30. In the case of FCI, we also show the first excited state energy. The CCSD solution corresponds the positive excitation energy solution and the negative energy solution is not shown. \textbf{B}.~The same data visualized in terms of the angle $\phi$ in the $gh$ plane, as defined using polar coordinates. \textbf{C}.~The largest cluster amplitude in $T_1$. }
    \label{fig:heh2-loop-mpn}
\end{figure*}

\section{Møller-Plesset perturbation theory}
\label{Sec:Benchmark}
\noindent

\revE{\revHK{We now} consider}
Møller-Plesset perturbation theory,\citep{helgaker2013molecular} 
\revE{where}
the ground state wave function is similarly expressed using intermediate normalization with respect to a reference wave function. If this reference is $\revE{\ket{\Phi} = \ket{\mathrm{HF}}}$, then the wave function is given as
\begin{equation}
\begin{aligned}
    \ket{\Psi} = \ket{\mathrm{MP}} = \ket{\mathrm{HF}} + \ket{\mathrm{MP^{(1)}}} + \ket{\mathrm{MP^{(2)}}} + \ldots ,
\end{aligned}
\end{equation}
where 
\begin{align}
\Dbraket{\mathrm{HF}}{\mathrm{MP}} = 1 
\end{align}
and the superscripts denote orders in the fluctuation potential.
Considering once again the \revHK{critical} point where the weight along the reference is zero, we see  that the zero'th order wave function $\ket{\mathrm{HF}}$ is not a component of the ground state wave function. \revHK{This implies, given the perturbation series is convergent and converges to an eigenstate of the Hamiltonian, that this must necessarily be an excited state, as the state is intermediately normalized. The same argument applies to the excited state that together with the ground state gives rise to the geometric phase effect. For the excited state, $\ket{\mathrm{R}_1}$ is used as a reference for the perturbation expansion. We can now explicitly state that
\begin{equation}
    \begin{aligned}
        \ket{\Psi_0} &=  \ket{\mathrm{HF}} +  \ket{\mathrm{MP}_0} \\
        \ket{\Psi_1} &=  \ket{\mathrm{R}_1} + \ket{\mathrm{MP}_1}
    \end{aligned}
    \label{eq:perturbed}
\end{equation}
are exact eigenfunctions for the ground state and the excite state, respectively, when away from the critical point. In Eq.(\ref{eq:perturbed}), the perturbation corrections are denoted $\ket{\mathrm{MP}_n}$. At the critical point, the two eigenfunctions $\ket{\Psi_0^\prime}$ and $\ket{\Psi_1^\prime}$ are orthogonal because the associated eigenvalues are assumed to be non-degenerate. As $\braket{\mathrm{HF}|\Psi_0^\prime}=1$ and $\ket{\Psi_0^\prime}$ is an eigenstate, $\ket{\Psi_0^\prime}$ cannot be the true ground state wave function, leaving only the excited state as possible. In the excited state the $\ket{\mathrm{R}_1}$ component must vanish and we have that
\begin{equation}
        \braket{\mathrm{R}_1|\Psi_0^\prime} =   \braket{\mathrm{R}_1|\mathrm{MP}_0^\prime} = 0.
\end{equation}
A similar argument for the excited state implies that $\ket{\Psi_1^\prime}$ must be the ground state and
\begin{equation}
        \braket{\mathrm{HF}|\Psi_1^\prime} =   \braket{\mathrm{HF}|\mathrm{MP}_1^\prime} = 0.
\end{equation}
This shows that at the critical point, both perturbative corrections are orthogonal to both reference functions, and $ \braket{\mathrm{MP}_0^\prime|\mathrm{MP}_1^\prime} = 0$. \revN{In the above, we have used the fact that the critical point for the two states must necessarily occur at the same location on the loop. To see this, suppose that $\ket{\Psi_0^\prime}$ corresponds to the excited state at the ground-state critical point. If the critical point for the excited state were instead located elsewhere, then $\ket{\Psi_1^\prime}$ would also represent the excited state—contradicting the assumption that the two states are orthogonal. It follows that the critical points must coincide.
 } 

We can conclude that in a $(N-1)$ dimensional nuclear configuration space around the conical intersection seam, the roles of $\ket{\Psi_0}$ and $\ket{\Psi_1}$ are interchanged, and the ground state wave function represent the excited state and vice versa. When the perturbation expansion is
truncated, as in practical calculations, the energies are continuous functions of the nuclear coordinates. Consequently, the ground state potential energy curve will gradually change to the excited state curve and vice versa for the excited state, leading to two crossing points along the loop. In an extended region, the wave functions \emph{neither} represent the ground state nor the first excited state in any meaningful sense.
}

In Figure~\ref{fig:heh2-loop-mpn} we illustrate this behavior for a ground state conical intersection in \ce{HeH2}.\citep{gulania2021equation} We compare, along a loop around the intersection, the Møller-Plesset potential energy at orders $n = 2,4,10$, $20$, $30$ with CCSD and with exact full configuration interaction (FCI). The CCSD and FCI energies were obtained using the $e^T$ \revHK{program}\citep{folkestad20201} and \revE{the} Møller-Plesset energies using Psi4.~\citep{smith2020psi4} 

\begin{figure*}[htb]
    \centering
    \includegraphics[width=\linewidth]{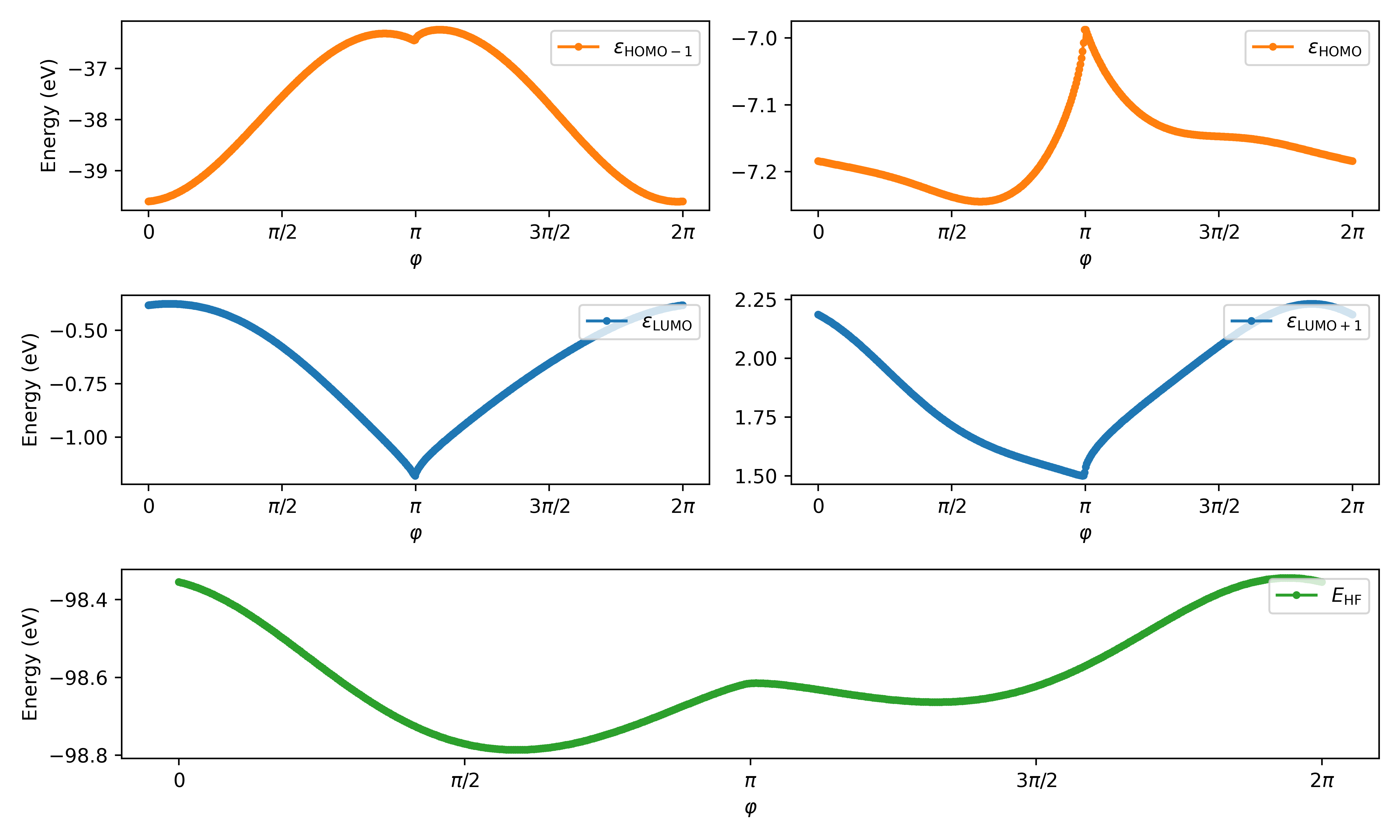}
    \caption{Selection of orbital energies and Hartree-Fock energy for the loop in the branching plane about the \ce{HeH2}/aug-cc-pVDZ intersection.}
    \label{fig:orben}
\end{figure*}

\revE{Somewhat surprisingly}, the observed distortions of the potential energy surfaces \revE{actually} become more pronounced at higher orders, where the Møller-Plesset potential energy surface sharply transitions from representing the ground state to representing the first excited state (around $\varphi = \pi$). This produces \revE{a large} unphysical artificial potential energy barrier \revE{(of 1.0 eV)} and cusp in the ground state potential energy surface. Inspecting the orbitals, we can trace the issue to the appearance of cusps in the orbital energies \revE{(see Figure \ref{fig:orben} for a selection of orbital energies)}. Note that the cusp in the Hartree-Fock energy is not so pronounced, in contrast to the Møller-Plesset energy. The non-analytical behaviour of the Hartree-Fock model has \revHK{previously} been analyzed by \v{C}\'{\i}\v{z}ek and Paldus for a model Hamiltonian.\cite{cizek1971}
 \revHK{The simple example presented here suggests that these regions are widespread around ground state conical intersections or close to areas where Hartree-Fock has bifurcations.}

\revE{\revHK{In general, }electronic structure} methods that depend on the Møller-Plesset ground state will also be affected by these issues. \revHK{An} example is the algebraic diagrammatic construction \revE{(ADC)} method for excited states,\citep{dreuw2023algebraic} where the excited state \revE{potential energy surfaces} would exhibit \revE{similar} artifacts \revE{to those found for the Møller-Plesset method} (including the cusp) due to the ground state MP$n$ energy contribution \revE{in ADC}. 

\revHK{Complete active space perturbation theory (CASPT)~\citep{Roos1990,helgaker2013molecular} is another method that can show the same behavior as Møller-Plesset perturbation theory and converge to an excited state in regions where the vanishing component theorem dictates \revE{that} the CAS reference's contribution to the exact wave function \revE{vanishes.} Artifacts \revE{close to conical intersections} in CASPT2 and multistate CASPT2 have been observed and discussed by Serrano-Andrés \textit{et. al.}~\cite{Serrano2005} 
For example, they investigated the avoided crossing in \ce{LiF}, where the CASPT2 ground state crosses the excited state in two places, indicative of the ground state converging to the excited state and vice versa\revE{, although this would not, in this case, be caused by a phase effect, since such  effects do not occur for diatomics due to the non-crossing rule.} 
In a recent review, Battaglia \textit{et. al.}~\cite{Battaglia2023}  discuss different variants of the CASPT2 method, and it would be interesting to analyze these in light of the vanishing component theorem; however, this requires a computational study that goes beyond the scope of the current article.}

\section{Conclusions}
\label{Sec:Conclusions}
\noindent
In this work, we have investigated the impact of the geometric phase effect on the description of ground state intersections in electronic structure methods. In particular, we have shown that enforcing a constant value on a wave function component (intermediate normalization) is incompatible with a correct description of the geometric phase effect, \revE{in the sense that it will} induce an asymptotic discontinuity along loops in internal coordinate space that \revE{enclose} an intersection. For some methods, this \revE{produces} complete breakdowns along paths that \revE{enclose} a ground state conical intersection. 

\revE{We have shown} how such breakdowns can occur in 
coupled cluster and Møller-Plesset methods in regions where the weight along the reference state vanishes \revE{due to the geometric phase effect}. This \revE{result} is especially worrisome, as the geometric phase effect is fundamentally a global effect \revE{in terms of changes in internal coordinates}: \revE{the path} around the intersection \revE{is} arbitrary\citep{longuet1975intersection} and any path enclosing an odd number of intersections will have a point where \revE{these methods break} down. Although it remains unknown how widespread these failures are, they may point to a need to reconsider the parametrization of the ground state in multi- and single-reference theories. 

One \revHK{promising} approach is to effectively remove conical intersections with the coupled cluster ground state by projecting the first excited state out of the cluster amplitudes and subsequently diagonalizing the \revE{similarity transformed} Hamiltonian to obtain the electronic states \revE{and the associated energies}. 
\revE{As is well-known, conical intersections are correctly described so long as the states are determined simultaneously by diagonalizing an effective Hermitian Hamiltonian. However, the fact that the geometric phase effect leads to breakdowns suggests that the removal of phase-inducing contributions (by projection onto the complement of the first excited state) provides a simple recipe to correct the unphysical description of ground state intersections found in some electronic structure methods.}
In a recent article, we show that this approach leads to a correct description of ground state conical intersections within a modified coupled cluster framework, and that it can be \revE{generalized to} other electronic structure methods that cannot properly describe intersections with the ground state.~\citep{rossi2024generalized} \revHK{One of us has recently developed the convex Hartree-Fock (CVX-HF) method,\citep{Rossi2025cvx} \revE{which was} proposed in Ref.~\citenum{rossi2024generalized}. This method can describe ground state conical intersections and avoids the non-analytical behavior discussed above. }

\begin{acknowledgments}
\noindent
\revE{We thank Sara Angelico for providing the ethylene MECI structure and the $\boldsymbol{g}$ and $\boldsymbol{h}$ vectors.}
We also thank Federico Rossi and Sara Angelico for fruitful discussions. We acknowledge funding from the European Research Council (ERC) under the European Union's Horizon 2020 Research and Innovation Programme (grant agreement No. 101020016).

\end{acknowledgments}

\bibliography{bibliography}

\clearpage

%
\end{document}